# Design of the Helium Purifier for IHEP-ADS Helium Purification System

Zhang Jianqin[1, 2;1], Li Shaopeng[1], Zhang Zhuo[1], Ge Rui[1]

[1]Institute of High Energy Physics, Chinese Academy of Science, Beijing, 10049, China
[2] University of Chinese Academy of Sciences, Beijing 100049, China

**Abstract**: Helium Purification System is an important sub-system in the Accelerator Driven Sub-critical System of the Institute of High Energy Physics (IHEP-ADS). The purifier is designed to work at the temperature of 77K. The purifier will work in a flow rate of 5g/s at 20MPa in continuous operation of 12 hours．The oil and moisture are removed by coalescing filters and a dryer, while nitrogen and oxygen are condensed by a phase separator and then adsorbed in several activated carbon adsorption cylinders. After purification, the purified helium has an impurity content of less than 5ppm.

摘要：不纯氦气回收净化系统是高能所 ADS 加速器次临界系统中的一个重要的子系统。纯化器的设计工作温度为 77 K，纯化器的在压力为 20MPa 和流量为 5g/s 条件下能连续工作 12 小时。氦气杂质中的油和水分别通过碳分子筛和干燥器去除，氮气和氧气通过液空分离器和活性炭吸附柱来去除。经过纯化后，纯净氦气中杂质含量小于 5ppm。

## 1. Introduction

There are a lot of superconducting facilities in the Accelerator Driven Sub-critical System of the Institute of High Energy Physics (IHEP-ADS), which need an enlarged cryogenic system to supply more liquid helium. Since helium is rare and expensive, the helium purification system is under construction to meet the high purity standard of the cryogenic system. The impure helium is mainly from the tests of superconducting cavities and the failures of the cryogenic system. The impurities of the impure helium can solidify at low temperature, which will choke the tubes and damage the turbo expanders. The helium purification system can keep the purity of helium for 99.9995% and has a storage capacity of 20000Nm$^3$ (including impure helium and pure helium) at 20MPa. The simplified process scheme of the purification system is in Fig.1, which includes gas bag, compressor, coalescing filters, dryer, helium purifier and high pressure cylinder manifold.

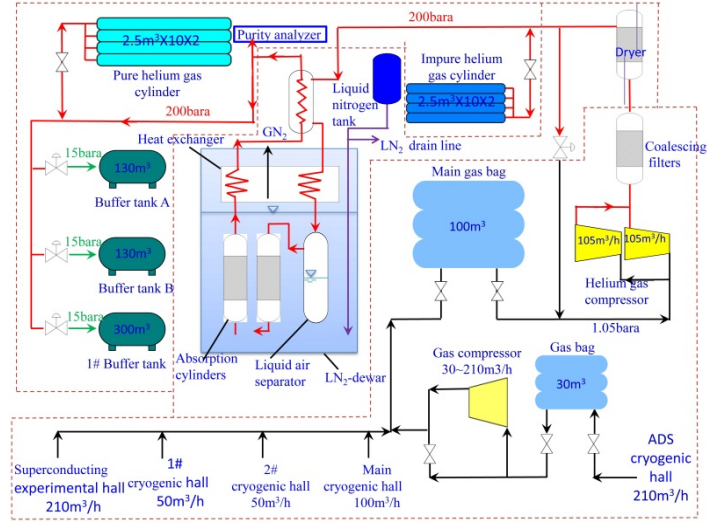

Fig.1 Simplified scheme of the purification system

Helium purifier is a key equipment of the helium purification system. Some countries have built their own high pressure purifiers [1, 2]. The design parameters of the helium purifier of IHEP-ADS are in table 1. The purifier works under the condition of 77K, 20MPa, which mainly includes high pressure heat exchanger, liquid air separator and carbon activated adsorption cylinders. In this work, the working principle and the design of the helium purifier are presented, and the studies of heat exchanger and the adsorbents are carried out.

Table 1. The design parameters of the helium purifier

| Pressure MPa, | Temperature K | Flow rate g/s | Run time h | Input gas purity % | Output gas purity % |
|---|---|---|---|---|---|
| 20 | 77 | 5 | 12 | 95 | 99.9995 |

## 2. Working principles of the helium purification

The impurities of the impure gas mainly include nitrogen and oxygen, with a minority of moisture, carbon dioxide, hydrogen and oil. Firstly, the impure helium will be gathered in the helium gas bag. When the gas bag reaches a high level, the impure helium will be compressed to the high pressure cylinder manifold. As the impure helium reaches a certain amount, the purifier begins to operate continuously with a throughput of 5g/s at 77K, 20MPa. The impure helium flow to the gas bag and compressed by the compressor, then passes through the oil and water separator vessel, coalescing filters, dryer, outer heat exchanger, high pressure heat exchanger, condenser, liquid air separator and adsorption cylinders. At last, the pure helium will be stored in the high pressure cylinder manifold. Most of the nitrogen and oxygen are liquefied in the liquid air separator and the rest are adsorbed by the adsorbents. The condensate is blown down about 3 hours periodically. The purifier is designed for an uptime of 12 hours, the impurities at the outlet are detected by the analyzer. The regeneration starts by any one of the following three criterions:

- the time in excess of the uptime
- the impurity at the outlet in excess of 5ppm
- manual triggering of the regeneration

For the regeneration, liquid nitrogen is drained out and adsorption cylinders undergo regeneration by heating and evacuation. Firstly, the purifier is depressurized and condenser is emptied. Then the adsorbent cylinders are heated up. The liquid nitrogen in the dewar is emptied to the buffer tank when the pressure of the dewar increases. The adsorbent cylinders are heat to 100℃ with an evacuation . The regeneration is completed by pressurization with pure helium. The purifier is equipped with an automatically process control system and a local operator panel. So the purification and the regeneration process can be controlled automatically and manually.

## 3. Design of helium purifier

The purifier includes dewar, outer heat exchanger, high pressure heat exchanger, condenser, liquid air separator and adsorption cylinders. The dewar is fitted with a quantity of liquid nitrogen, in which condenser, liquid air separator and adsorbent cylinders are submerged. The height of the purifier is 3050mm, and the diameter of dewar is 600mm with a thickness of 5mm. The outer heat exchanger acts as dryer which brings down the water dew point to approximately 1℃. The helically coiled tube-in-tube heat exchanger is selected as the high pressure heat exchanger, which is compact and high efficient. The outside diameter of the inner tube is 8mm with a wall thickness of 1.5mm, while the outside diameter of the outer tube is 16mm with a wall thickness of 3mm. The diameter of the coil is 270mm. In order to decrease the drop pressure of the tube, the heat exchanger is divided into five layers. The result of the high pressure heat exchanger is shown in table 2 and the tube length is 11.84m. The liquid air separator (vol. 18.3l) collects the liquid air at the bottom of the separator. After the liquid air separator, the impurities of the gas are less than 0.32%. There are 6 adsorption cylinders (vol. 18.3l) which are divided into two ways and connected in series. The adsorbent in the adsorber is coconut shell activated carbon with a good performance at 77K.

Table 2. Parameters of the helically coiled tube-in-tube heat exchanger

|  | Input temperature K | Input Pressure MPa | Flow rate g/s | Output temperature K | Pressure drop Pa |
|---|---|---|---|---|---|
| **Input gas** | 300 | 20 | 5 | 93.5 | 5140 |
| **Output gas** | 78 | 19.8 | 4.9 | 295 | 2360 |

### 3.1 Numerical studies of high pressure heat exchanger

The designed helically coiled tube-in-tube heat exchanger is model with CFD methods [3, 4]. The input gas is in the annulus and the output gas is in the inner tube. These two fluids are counter-flow. Fluent 14.0 in ANSYS is used to analysis the heat transfer and fluid flow in the heat exchanger. The transport and thermal properties of helium is dependent on the temperature. The RNG k-ε model with standard wall functions is used in the analysis. The boundary condition of inlet is velocity, as the outlet is pressure. The convergence residual is 1.0e-05. The results show that the difference of the outlet temperature between the calculated and the numerical is less than 0.5% and the difference of the pressure drop is less than 20%. The temperature distribution of the whole heat exchanger is shown in Fig.2. And the sectional views of the outlet temperature in the inner tube and the annulus are shown in Fig.3. The secondary flow caused by the centrifugal forces can enhance the heat transfer in the inner tube and the annulus.

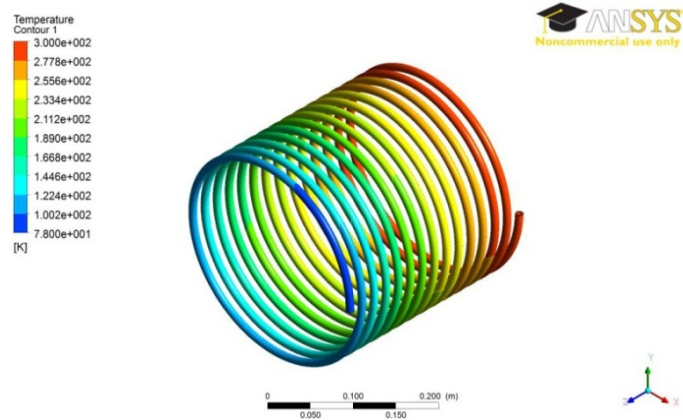

Fig.2. Temperature profiles of the helically coiled tube-in-tube heat exchanger

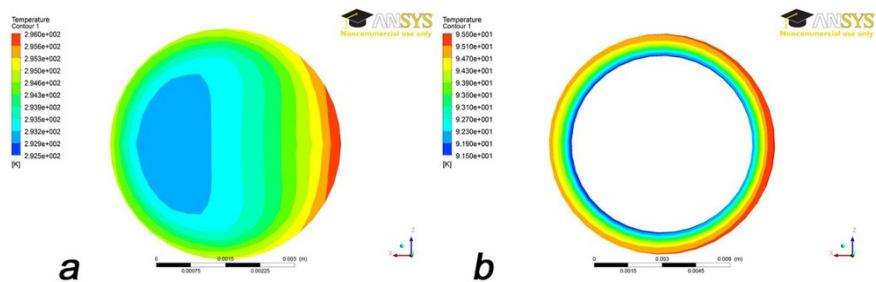

Fig.3. Temperature profiles of the outlet in (a) the inner tube and (b) the annulus

### 3.2 Experimental studies of the adsorbents

The Linde activated carbon is widely used in the purification. The properties of the static and dynamic adsorption of Linde coconut shell activated carbon are studied experimentally. The adsorption isotherms of $N_2$ and $O_2$ (77K) are presented in Fig.4a. In the figure, it shows that the static adsorption capacity of $O_2$ is larger than that of $N_2$ and they belong to type I. The specific surface area is 999.72$m^2$/g which is fitted with $N_2$ adsorption isotherm by the BET model [5]. The pore size distribution is analyzed using Density Functional Theory (DFT) method with $N_2$ adsorption isotherm, which is shown in Fig.4b. The total pore volume of Linde activated carbon is 0.465$cm^3$/g and micropore occupies approximately 95% of the total volume.

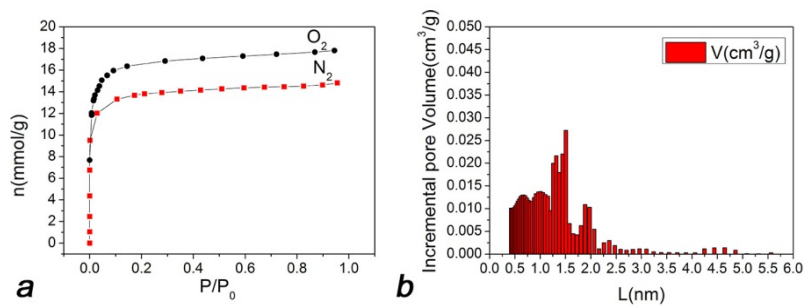

Fig.4. (a) N2 and O2 static adsorption isotherms (77K); (b) Pore size distribution of Linde activated carbons

The breakthrough curves of $N_2$ and $O_2$ are shown in Fig.5a, which is for the adsorption of a binary mixture of $O_2$ and $N_2$ (22:78) on the adsorbent of the Linde activated carbon. In Fig.5a, $N_2$ breaks through firstly. Then the concentration of $N_2$ in the outlet increases and reaches a maximum, which is higher than the concentration of $N_2$ in the entering gas stream. The concentration of $O_2$ leaving the bed increases, with that of $N_2$ decreasing. Finally, the concentrations of $N_2$ and $O_2$ in the leaving gas and the entering gas are same. The breakthrough adsorption capacity is the adsorption capacity of the component when the component breaks through. Fig.5b presents the breakthrough adsorption capacity in a binary mixture and the static adsorption capacity of $N_2$ at different partial pressure. When $P/P_0>0.1$, the $N_2$ breakthrough adsorption capacity has little changes. The breakthrough adsorption capacity is approximately 44% less than the static adsorption capacity. The amount of the activated carbon can be calculated by the $N_2$ static adsorption capacity with a factor of 1.5-2.

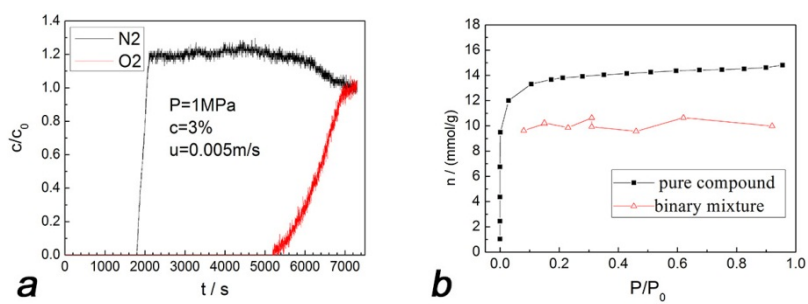

Fig.5. (a) Breakthrough curves of a binary mixture (O2 and N2); (b) N2 breakthrough adsorption capacity vs. N2 static adsorption capacity

## 4. Conclusion

High pressure purifier is a key equipment of the helium purification system. The working principles of the helium purification and regeneration are described. The design of the purifier is shown in details. It's known that the high pressure heat exchanger and the adsorption cylinders are important parts of the purifier. The helically coiled tube-in-tube heat exchanger is numerically studied and can meet the need of heat transfer. The experiments of static and dynamic adsorption of Linde coconut shell activated carbon are carried out. The activated carbon has 95% micropore which can adsorb the impurities well at low pressure. The amount of the activated carbon can be calculated by the $N_2$ static adsorption capacity with a factor of 1.5~2.